\documentclass[aps,prb,twocolumn,groupedaddress,showpacs]{revtex4}

\usepackage[bookmarks]{hyperref}
\usepackage[dvips]{graphicx}
\usepackage{amsmath}
\usepackage{amstext}
\usepackage{graphics}
\usepackage{exscale}
\usepackage{epsfig}
\usepackage[T1]{fontenc}
\usepackage{bm}
\usepackage{bbm}

\usepackage{latexsym}

\renewcommand{\epsilon}{\varepsilon}

\newcommand{\unitop}{\mathbbm{1}}

\newcommand{\partdera}[2]{\frac{\partial #2}{\partial #1}}

\newcommand{\expval}[2]{ \langle  #1 #2\ \!\! \rangle}

\newcommand{\elcre}[2]{ c^{\dagger}_{#1,#2}}
\newcommand{\elann}[2]{ c_{#1,#2}}

\newcommand{\vk}{{\bm k}}
\newcommand{\vq}{{\bm q}}

\newcommand{\Hand}{H_{\mathrm{And}}}
\newcommand{\thGf}{{\cal G}}

\newcommand{\Imag}{\mathrm{Im}}
\newcommand{\Real}{\mathrm{Re}}

\newcommand{\hc}{\mathrm{h.c.}}

\begin{document}

\title{Quasiparticle excitations and dynamic susceptibilities in the BCS-BEC crossover} 
\author{J. Bauer${}^1$ and A.C. Hewson${}^2$}
\affiliation{${}^1$Max-Planck Institute for Solid State Research, Heisenbergstr.1,
  70569 Stuttgart, Germany}
\affiliation{${}^2$Department of Mathematics, Imperial College, London SW7 2AZ,
  United Kingdom}
\date{\today} 
\begin{abstract}
We study dynamic ground state properties in the crossover from weak (BCS) to strong
coupling (BEC) superfluidity.  Our approach is based on the attractive Hubbard
model which is analyzed by the dynamical mean field theory (DMFT) combined
with the  numerical renormalization group (NRG). We present an extension of
the NRG method for effective impurity models to self-consistent
calculations with superconducting symmetry  breaking.  In the one particle
spectra we show quantitatively how the Bogoliubov quasiparticles at weak
coupling become suppressed at intermediate coupling. We also present results for
the spin and charge gap. The extension of the NRG method to self-consistent
superconducting solutions opens the possibility to study a range of other
important applications.     
 \end{abstract}
\pacs{71.10.Fd,73.22.Gk,74.20.Fg,71.27.+a}


\maketitle

\paragraph*{Introduction -}
There has been a resurgence of interest into the nature of the
 crossover from the weak coupling \citet*{BCS57} (BCS)
superfluidity to the Bose Einstein condensation (BEC) of preformed pairs, due
 to its recent experimental realization for ultracold
atoms in an optical trap\cite{GRJ03,ZSSRKK04,ZSSSK05,BDZ08}. In the ultracold
atom experiments, the interactions between the fermionic atoms in the  trap
can be tuned by a Feshbach resonance   permitting an unprecedented control of
the parameters, so that theoretical predictions may now be put to an
experimental test.  Above the transition temperature the two limiting cases,
BCS superconductivity and BEC, correspond to quite different physical
situations. The weak coupling BCS theory describes a fermionic system with
Cooper instability, whereas the strong coupling BEC limit is a
system of strongly bound fermion pairs, which obey Bose statistics.  The theoretical
understanding which has been developed  over the years is that the properties,
such as the order parameter $\Delta_{\rm sc}$ and 
the transition temperature $T_c$ to the superfluid state, are connected by a smooth
crossover, and approximate interpolation schemes between these limits have been devised
 \cite{Eag69,Leg80,NS85,Ran95}.\par

There is experimental evidence that the  BCS-BEC crossover has also relevance
for strong coupling and high temperature superconductors. These
superconductors display  properties, it has been
claimed \cite{MRR90,TCC05,CLS06}, that can be  understood in terms of
local pairs, preformed  above the transition temperature $T_c$, in contrast to the
BCS picture, where the pairs no longer exist above $T_c$. \par 

As more experimental techniques, such as pairing gap spectroscopy \cite{CBARJDG04,GRJ05},
are being developed to probe this crossover, more detailed theoretical
predictions are required. In particular the dynamic response functions have
received little theoretical attention, and it is the predictions for these
quantities through the crossover that will be the focus of the 
present paper.

\par

Here, we use the attractive Hubbard model \cite{MRR90} to study the
crossover. In order to investigate the full crossover regime from weak to 
strong coupling a reliable non-perturbative method is needed. We employ the
dynamical mean field theory (DMFT), together with the numerical
renormalization group (NRG) to treat the effective impurity problem. DMFT
studies, which are appropriate for the three and higher dimensional case, for
the attractive Hubbard model with other methods for the effective 
impurity problem  have been carried out in the normal phase\cite{KMS01,CCG02},
and more recently in the symmetry-broken phase\cite{GKR05,TBCC05,TCC05}. But spectral
properties have not been calculated and analyzed in detail there. There have
been  calculations of the dynamic response functions for the two dimensional
model, for example with quantum Monte Carlo \cite{SPSBM96} or recent work using
the cellular DMFT \cite{KGT06}. Our results based on the DMFT are not expected
to be directly applicable to this low dimensional case, but a comparison can
be of interest to see if there are common trends. \par 

Our method to study the effective impurity problem, the NRG,
is a non-perturbative method originally devised by Wilson\cite{Wil75} to
analyze the many body problem in the Kondo model and Anderson impurity model
(AIM). In this approach the high energy
degrees of freedom are eliminated numerically  to find a an effective low
energy theory, and its early  application contributed vitally to the 
quantitative understanding of the Kondo problem \cite{hewson}. 
The method was developed further in the 1990s, when it was extended to the
calculation of spectral functions \cite{SSK89,CHZ94}. After the advent of the
DMFT\cite{GKKR96} it was realized then that it could be 
used for self-consistent solutions of the effective impurity problem
\cite{BPH97}. 
The calculation of spectral functions was improved further by approaches based
on the density matrix \cite{Hof00} and the complete basis set proposed by
Anders and Schiller \cite{AS05,PPA06,WD07}.  
Our approach uses this latest full density matrix scheme to calculate reliable
spectral functions. The NRG method and its applications are  comprehensively
reviewed by Bulla et al. \cite{BCP08}.

In the original theory and in many of the applications one deals with a metallic
system, where the NRG inherent logarithmic discretization and the descent to
low energies gives the accurate low energy properties.
However, the method has also given reliable results in situations with a gap in the
spectrum. Examples include the Mott transition \cite{Bul99} or antiferromagnetic
solutions \cite{ZPB02,BH07c} in the repulsive Hubbard model in the DMFT-NRG
framework. There is also an 
extensive amount of work for situations where the bath of the impurity model
is a mean field BCS superconductor with given  energy gap
\cite{SSSS92,SSSS93,YO00,CLKB04,OTH04,BOH07,HWDB08}. The NRG parameters for the bath in
this case are fixed and can be determined by a straightforward generalization
of the original NRG approach. Here, we present the extension of the method to
calculations where the parameters describing the superconducting bath have to
be determined self-consistently, which is a requirement of the DMFT approach,
where the impurity is an effective one and plays an auxiliary role. This opens
the possibility to study a number of interesting non-perturbative problems
involving superconductivity in a well controlled framework.

\paragraph*{Formalism -}
Our study is based on the attractive Hubbard model \cite{MRR90},
\begin{equation}
H=-\sum_{i,j,\sigma}(t_{ij}\elcre {i}{\sigma}\elann
{j}{\sigma}+\hc)-\mu\sum_{i\sigma}n_{i\sigma}-U\sum_in_{i,\uparrow}n_{i,\downarrow},
\label{attHub}
\end{equation}
with the chemical potential $\mu$, the interaction strength $U>0$ and  the
hopping parameters $t_{ij}$. $\elcre {i}{\sigma}$ creates a fermion at site
$i$ with spin $\sigma$, and $n_{i,\sigma}=\elcre {i}{\sigma}\elann {i}{\sigma}$. 
For the DMFT calculation we employ the Anderson impurity model in a superconducting medium as the
effective impurity model,
\begin{eqnarray}
\Hand&=&H_{\rm imp}+
\sum_{\vk,{\sigma}}\epsilon_{\vk}\elcre{\vk}{\sigma}\elann{\vk}{\sigma} 
+ \sum_{\vk,{\sigma}}V_{\vk}(\elcre{\vk}{\sigma}{d}_{\sigma} + \hc)
\nonumber \\
&&
-\sum_{\vk}\Delta_{\vk}[\elcre{\vk}{\uparrow} 
\elcre{-\vk}{\downarrow}+\elann{-\vk}{\downarrow}\elann{\vk}{\uparrow}],
\label{scAIMn}
\end{eqnarray}
where $H_{\rm
  imp}=\sum_{\sigma}\epsilon_dn_{d,\sigma}-Un_{d,\uparrow}n_{d,\downarrow}$
and $d_{\sigma}$ is the fermionic operator at the impurity site.  
$\epsilon_{\vk}$, $V_{\vk}$ and  $\Delta_{\vk}$ are parameters of the
medium. The non-interacting  Green's function matrix at $T=0$ has the form,
\begin{equation}
 \underline{G}_0(\omega)^{-1}=
\omega\unitop_{2}-\epsilon_{d}\tau_3-\underline K(\omega).
\label{scG0}
\end{equation}
$\underline K(\omega)$ is the generalized matrix hybridization for the medium,
with diagonal part
\begin{equation}
  K_{11}(\omega)=\frac1N\sum_{\vk}V_{\vk}^2
\frac{\omega+\epsilon_{\vk}}{\omega^2-(\epsilon_{\vk}^2+\Delta_{\vk}^2)} 
\label{k11sc}
\end{equation}
and offdiagonal part,
\begin{equation}
  K_{21}(\omega)=\frac1N\sum_{\vk}V_{\vk}^2
\frac{\Delta_{\vk}}{\omega^2-(\epsilon_{\vk}^2+\Delta_{\vk}^2)} . 
\label{k21sc}
\end{equation}
In the DMFT with superconducting symmetry breaking the effective Weiss field
is a  $2\times2$ matrix  $\underline\thGf_{0}^{-1}(t)$\cite{GKKR96}.    The
DMFT self-consistency  equation in this case also is a matrix equation,  
\begin{equation}
  \underline\thGf_{0}^{-1}(\omega)=\underline G(\omega)^{-1}
  +\underline\Sigma(\omega), 
\label{scselfcon}
\end{equation}
with the $\vk$-independent self-energy $\underline\Sigma(\omega)$ and the
local lattice Green's function $\underline G(\omega)$.
We identify  as usual $\underline{G}_0(\omega)=\underline\thGf_{0}(\omega)$.  
Having calculated the local Green's function $\underline G$ and the
self-energy $\underline\Sigma(\omega)$, the self-consistency equation
(\ref{scselfcon}) determines the new Weiss field and medium as input for the
effective impurity problem.   
Once $K_{11}(\omega)$ and $K_{21}(\omega)$ are given by (\ref{scselfcon}), the
problem is to calculate the effective impurity model parameters and map
(\ref{scAIMn}) to the linear chain Hamiltonian, 
\begin{eqnarray}
  \Hand&=&H_{\rm imp}  + \sum_{{\sigma},n} \beta_{n}
(f^{\dagger}_{n,\sigma}f_{n+1,\sigma}+  \hc) \\
&&+\sum_{\sigma,n}\epsilon_{n}f^{\dagger}_{n,\sigma}
  f_{n,\sigma}-\sum_{n}\Delta_{n}(f^{\dagger}_{n,\uparrow}f^{\dagger}_{n,\downarrow}+\hc), 
\nonumber 
\end{eqnarray}
to which the iterative diagonalization of the NRG can be applied. 
$\beta_{n}$, $\epsilon_{n}$, and $\Delta_{n}$ are the parameters of the linear
chain model and $f^{\dagger}_{n,\sigma}$ the operators for the sites. The details
of how this can be achieved will be published elsewhere \cite{BHD09pre}.

\paragraph*{Results -}
Static and integrated quantities like the order parameter, the average pair
density $\expval{n_{\uparrow}n_{\downarrow}}{}$ or superfluid density $D_s$
have been discussed in other works \cite{GKR05,TCC05}. We will present our 
DMFT-NRG  results for these quantities in a  separate publication
\cite{BHD09pre}. Here we focus on dynamic response functions, which have received
little attention  so far. 

For numerical calculations within the DMFT-NRG approach we take the
semi-elliptical form of the Bethe lattice for the non-interacting 
density of states  $\rho_0(\epsilon)={2}\sqrt{D^2-\epsilon^2}/{\pi D^2}$,
where $2D$ is the band width with $D=2t$ for the Hubbard model. $t=1$ sets the
energy scale in the following. All the results presented here are for $T=0$
and the generic case of quarter filling, $n=1/2$.

The spectral gap and the quasiparticle excitations can be analyzed from the
one particle Green's function $G_{\vk}(\omega)$, which is the diagonal element
of the inverse of $\underline G_{\vk}(\omega)^{-1}$
\begin{equation}
  \underline G_{\vk}(\omega)^{-1}=
  \underline G_{\vk}^0(\omega)^{-1}-\underline\Sigma(\omega).
\label{scdyson}
\end{equation}
In the BCS limit the excitations can be described by
\begin{equation}
G^{\rm BCS}_{\vk}(\omega)=
\frac{u_{\vk}^2}{\omega-E^0_{\vk}}+\frac{v_{\vk}^2}{\omega+E^0_{\vk}} ,
\nonumber
\end{equation}
where $u_{\vk}^2=(1+(\xi_{\vk}+Un/2)/{E^0_{\vk}})/2$, $v_{\vk}^2=1-u_{\vk}^2$
with $\xi_{\vk}=\epsilon_{\vk}-\mu$. The mean field order parameter
$\Delta_{\rm   sc, MF}=U\expval{c_{0,\uparrow}c_{0,\downarrow}}{}_{\rm MF}$
and the chemical potential $\mu$ are determined from the gap and number
equation, respectively. In this approximation the two bands of quasiparticle
excitations are given by $\pm
E_{\vk}^0=\pm\sqrt{(\xi_{\vk}+Un/2)^2+\Delta_{\rm  sc, MF}^2}$, with weights 
$u_{\vk}^2$ for the positive  and $v_{\vk}^2$ for the negative excitations
with infinite lifetime. In the real interacting system this 
is obviously not the case and our calculation of the dynamic quantities allows
one to study this. 

The standard way to analyze the real quasiparticle excitations $E_{\vk}$ is to
use the many-body definition, $\Real\, G_{\vk}(\omega=E_{\vk})^{-1}=0$.
When the self-energy functions are 
calculated from DMFT this can be solved for a given $\epsilon_{\vk}$ as an
implicit equation. Due to the symmetries of the self-energy for a solution
$\omega=E_{\vk}$ also $\omega=-E_{\vk}$ is a solution of this equation. In
order to extract quasiparticle parameters in a simplified form, we expand
around these solutions.\cite{expansion} We define   
\begin{equation}
  z(E_{\vk})^{-1}=\frac1{2E_{\vk}}\partdera{\omega}{\Real G_{\vk}(\omega=E_{\vk})^{-1}}.
\end{equation}
In the vicinity of $\omega=E_{\vk}$ the Green's function then reads
\begin{equation}
  G_{\vk}(\omega)=\frac{u^2(E_{\vk})}
{z(E_{\vk})^{-1}(\omega-E_{\vk})+i\bar W(E_{\vk})}, 
\end{equation}
where we have set $\omega=E_{\vk}$ in the numerator,
$u^2(E_{\vk})=(E_{\vk}+\xi_{\vk}-\Sigma_{22}(E_{\vk}))/2E_{\vk}$, 
and  approximated the imaginary part by 
\begin{equation}
  \bar W(E_{\vk})=\frac{\Imag G_{\vk}(\omega=E_{\vk})^{-1}}{2E_{\vk}}
\end{equation}
independent of $\omega$. The main contribution for the spectral density near
$\omega=E_{\vk}$ is then  
\begin{equation}
  \rho_{\vk}(\omega)=u^2(E_{\vk})z(E_{\vk})\frac{W(E_{\vk})/\pi}{(\omega-E_{\vk})^2+
    W(E_{\vk})^2},
\label{qplorentz}
\end{equation}
where $W(E_{\vk})=z(E_{\vk}) \bar W(E_{\vk})$. If
$W(E_{\vk})$ is small then this is a well-defined Lorentz quasiparticle peak
with width $W(E_{\vk})$ and spectral weight $w_+(E_{\vk})=z(E_{\vk})u^2(E_{\vk})$. The
same analysis can be done near $\omega=-E_{\vk}$. With these quantities we can
analyze the question up to which interaction strength there are well-defined
fermionic (gapped) quasiparticles, i.e. which are not too broad. 

\begin{figure}[!htbp]
\centering
\includegraphics[width=0.45\textwidth]{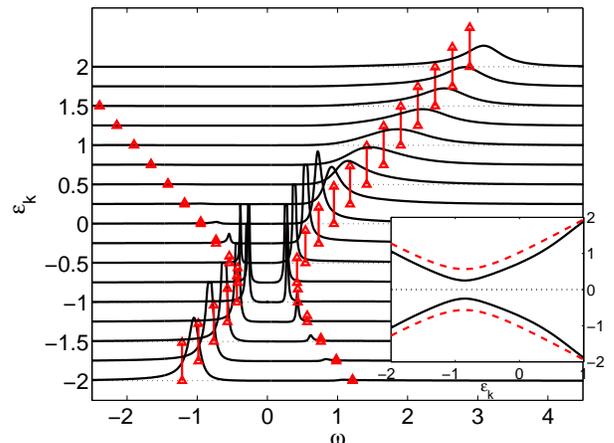}
\vspace*{-0.5cm}
\caption{(Color online) The $\epsilon_{\vk}$-resolved spectral functions
  $\rho_{\vk}(\omega)$ for $U=2$. The arrows
   show the delta-function peaks of the mean field result $\rho^{\rm
     BCS}_{\vk}(\omega)$, where the height of the arrow indicates the weight  
   of the peak. Inset: Mean field bands $E_{\vk}^0$ (dashed) and real
   quasiparticle band $E_{\vk}$  (full line). }  
\label{kresspecscU2}
\end{figure}

In Fig. \ref{kresspecscU2} we give a typical example of
$\epsilon_{\vk}$-resolved spectra $ \rho_{\vk}(\omega)$ for $U=2$, which is
the critical interaction $U_c$ for bound state formation in the two-body
problem for the Bethe lattice \cite{GKR05}. 

\noindent
We can see the two bands of quasiparticle excitations with the
sharpest peaks in the region of minimal spectral gap.
These spectra can be compared with the ones which were
calculated by Garg et al.\cite{GKR05} by iterated perturbation theory. There
the quasiparticle excitation delta peaks are disconnected from the continuum,
which is however an artifact of the approximation for the self-energy, whose
imaginary part vanishes over  too large a region in $\omega$. 
The mean field results (red arrows) describe the form of
the quasiparticle bands qualitatively well. Also the weight of the peaks in
the full spectrum $\rho_{\vk}(\omega)$ is comparable with the height of the
arrows of the mean field theory. Notice that the sharpest peaks of
$\rho_{\vk}(\omega)$ have the Lorentzian shape as in (\ref{qplorentz}). For
different $\epsilon_{\vk}$ and also at larger $U$ the peak form can be asymmetric
and (\ref{qplorentz}) not such a good fit.

\begin{figure}[!htbp]
\centering
\includegraphics[width=0.45\textwidth]{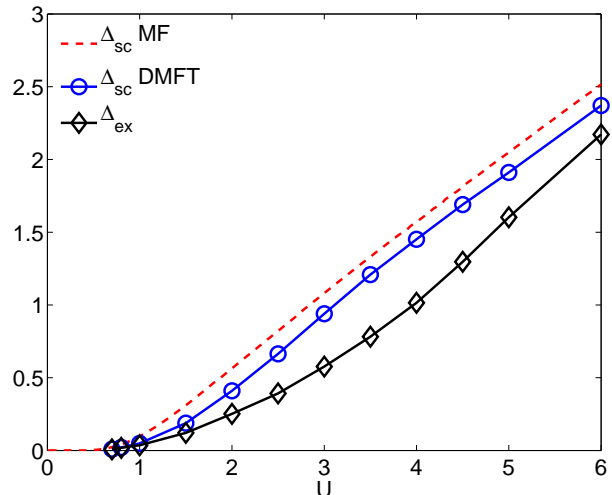}
\vspace*{-0.5cm}
\caption{(Color online) Comparison of the order parameter $\Delta_{\rm sc}$ (mean field theory
  and DMFT) with spectral gap $\Delta_{\rm ex}$ for a range of interactions $U$.}   
\label{gap_Udep_x1}
\end{figure}

The effect of the dynamic fluctuations on the excitation spectrum can be seen
clearer in the inset of Fig. \ref{kresspecscU2}, where we compare the
mean-field bands $E_{\vk}^0$ with the real quasiparticle band $E_{\vk}$. 
We can see a substantial reduction of the excitations energy for a certain bare
energy $\epsilon_{\vk}$. For instance, the minimal spectral gap $\Delta_{\rm
  ex}=\min(E_{\vk})$ is reduced by a factor of about $2.3$ in this case.
For different values of the interaction in the whole crossover regime  this
reduction effect can be seen in Fig. \ref{gap_Udep_x1}, where we plot the value of
the order parameter  computed by mean field (MF) theory, $\Delta_{\rm  sc,MF}$,
and DMFT, $\Delta_{\rm  sc, DMFT}$, and the spectral excitation gap $\Delta_{\rm
  ex}$. Note that only at weak coupling the minimal spectral gap is
strictly given by $\Delta_{\rm ex}$, since at intermediate to stronger
coupling there are broader peaks and one finds finite weight for
$|\omega|<\Delta_{\rm ex}$ implying a smaller excitation gap.\cite{BHD09pre}

\noindent
The order parameter $\Delta_{\rm  sc, DMFT}$ is reduced from its mean field
value $\Delta_{\rm  sc, MF}$ upon including fluctuations  for all
coupling strengths.\cite{MF92} Furthermore, we 
can see that the gap in the single particle spectrum  $\Delta_{\rm   ex}$, as
deduced from the one particle Green's function, is in turn smaller than
$\Delta_{\rm  sc, DMFT}$ for intermediate to strong coupling. At weak 
coupling these quantities tend to the same value suggesting an excitation
spectrum of a renormalized mean field form. 

\begin{figure}[!htbp]
\centering
\includegraphics[width=0.45\textwidth]{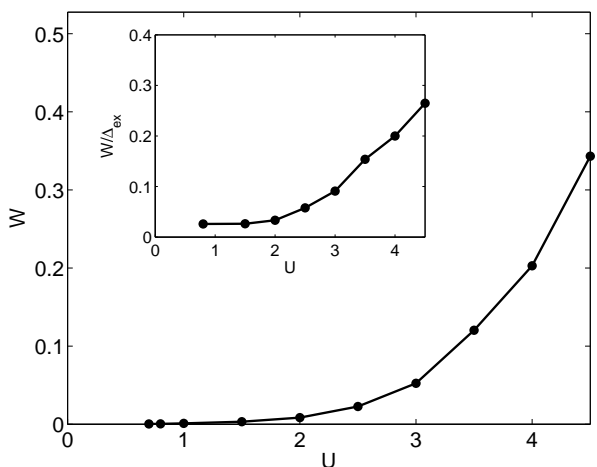}
\vspace*{-0.5cm}
\caption{The quasiparticle width $W$ at the excitation gap energy for a range
  of interactions $U$. Inset: The ratio $W/\Delta_{\rm ex}$.}    
\label{width_Udep_x1}
\end{figure}

In the weak coupling mean field theory the elementary excitations are gapped
fermions. However, for strong coupling we deal with a bosonic
system, and due to the large gap of order $U$ fermionic excitations are
suppressed. We now analyze at which point in the intermediate
coupling regime one tends to lose the sharp quasiparticle excitations. In
order to study this question we look at the width of the 
excitations defined at the smallest spectral excitations
$W=W(E_{\vk}=\Delta_{\rm ex})$. If this is small we have a strongly peaked 
well-defined quasiparticle excitation as observed in Fig. \ref{kresspecscU2},
whereas for larger width the excitation decays more rapidly. For different local
interaction strengths this is illustrated in Fig. \ref{width_Udep_x1}.   

\noindent
We find that whilst $W$ is very small for weak coupling it increases rapidly
in the intermediate coupling regime, $U=2-4$. The ratio $W/\Delta_{\rm ex}$,
which is plotted in the inset of Fig. \ref{width_Udep_x1}, also increases from a
small value at weak coupling to larger values at intermediate coupling. This  shows
that sharp (long-lived) quasiparticles excitations are suppressed then.

We now turn to spin and charge response functions.
Within our framework we can calculate the local dynamic spin susceptibility
$\chi_s(\omega)=(1/N)\sum_{\vq}\chi_s(\omega,\vq)$ and charge susceptibility
$\chi_c(\omega)$. The temperature dependence of $\chi_s$ in the static limit
has been discussed by Keller et al.\cite{KMS01} and the existence of a spin gap
was demonstrated.   In the following Fig. \ref{spinsusc} we show the
imaginary part of $\chi_s(\omega)$ (top) and $\chi_c(\omega)$ (bottom) for a
number of values of $U$.   

\begin{figure}[!htbp]
\centering
\includegraphics[width=0.45\textwidth]{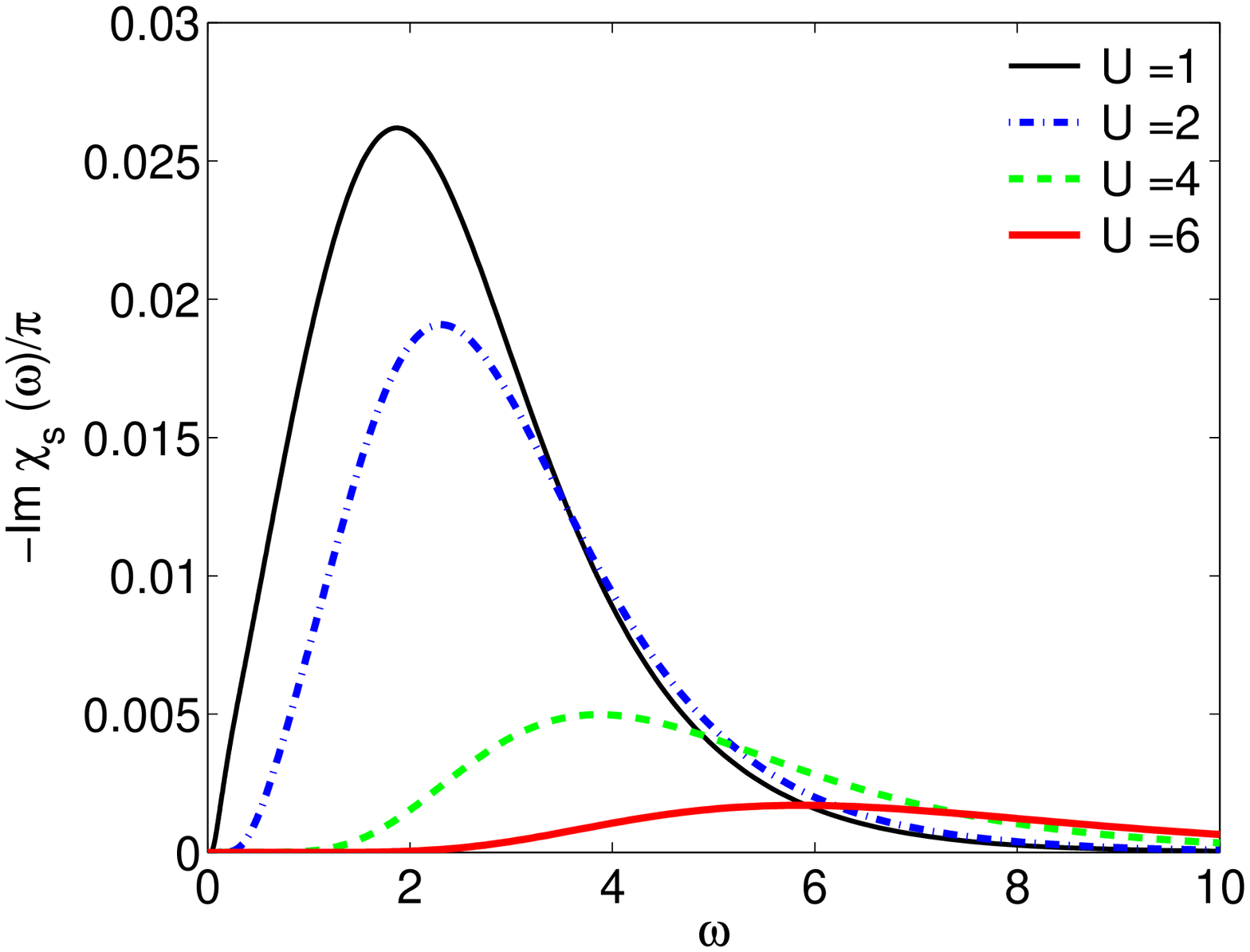}
\includegraphics[width=0.45\textwidth]{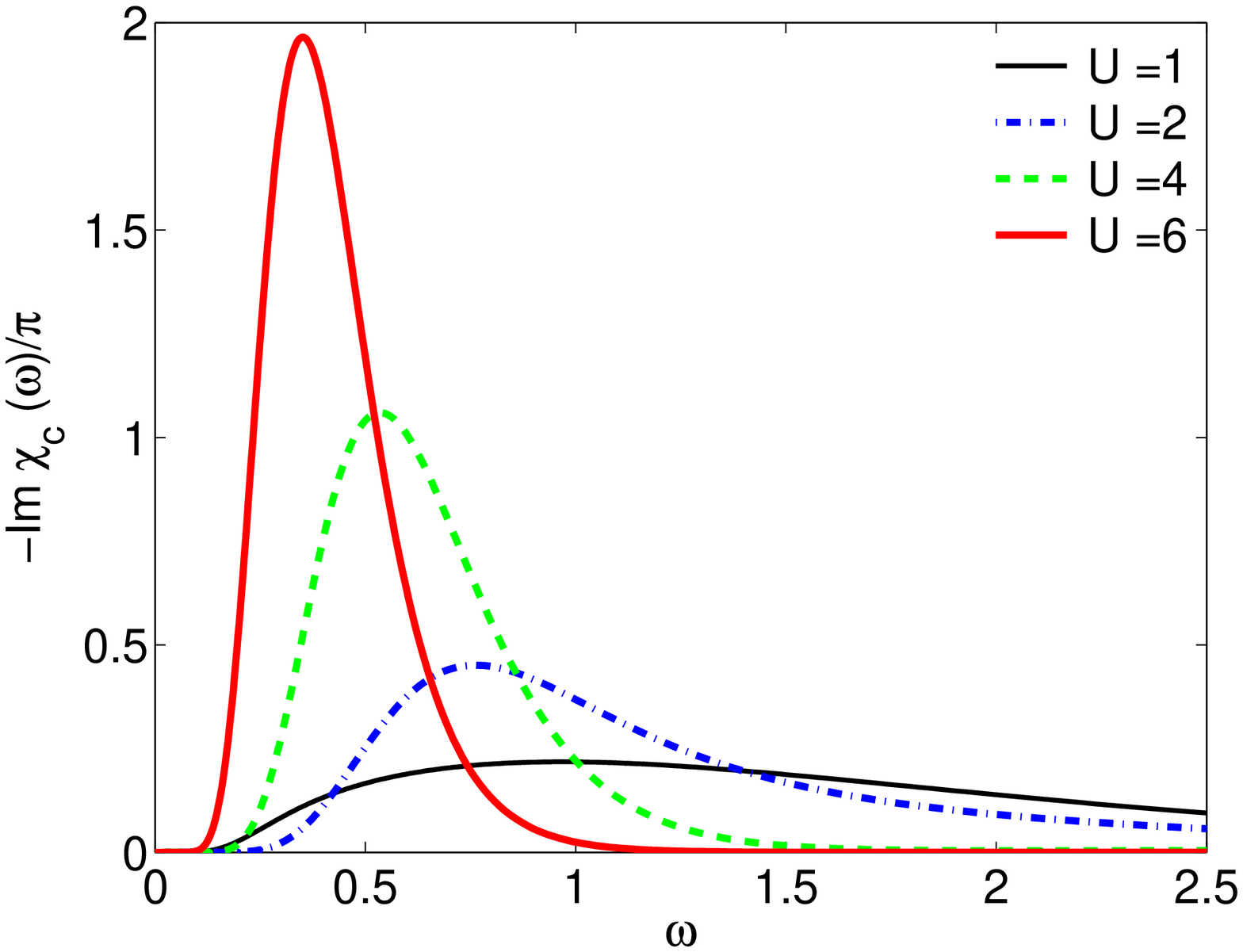}
\vspace*{-0.5cm}
\caption{(Color online) The imaginary part of the local dynamic spin susceptibility
  $\chi_s(\omega)$ (top) and of the charge susceptibility $\chi_c(\omega)$
  (bottom) for different $U$.}
\label{spinsusc}
\end{figure}
\begin{figure}[!htbp]
\centering
\includegraphics[width=0.45\textwidth]{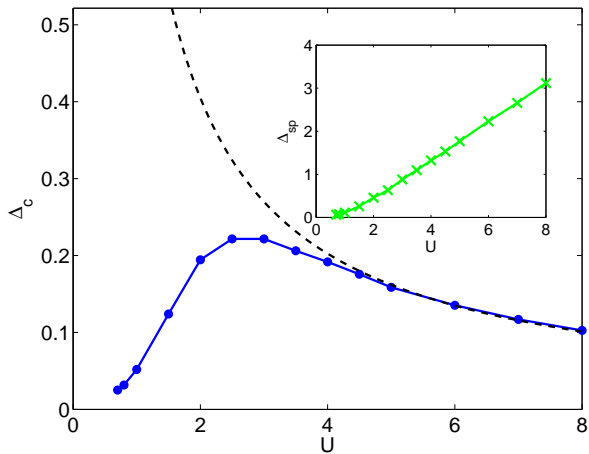}
\vspace*{-0.5cm}
\caption{(Color online) The charge gap $\Delta_{\rm c}$ for a range
  of interactions $U$. The dashed line behaves $\sim 4t^2/U$. Inset: The
  spin gap as a function of $U$.}    
\label{ch_gap_Udep_x0.5}
\end{figure}

\noindent
The spin response is suppressed when the local attraction $U$ is increased.
The spin gap\cite{extrchis} $\Delta_{\rm sp}$ grows with $U$ as can be seen in
the inset of Fig. \ref{ch_gap_Udep_x0.5}, and it is directly related to the
gap in the one particle  excitation spectrum. This is expected since spin excitations are
accompanied by breaking of a fermionic pair with opposite spin. For large $U$
the binding energy is proportional to $U$.  
The charge susceptibility shows a different behavior. The broad peak for weak
coupling becomes sharper in the strong coupling limit and moves to lower
energy. A similar trend is seen in the results for the charge susceptibility
for $\vq=0$ in the two dimensional model \cite{KGT06}. A peak related to the
massless Goldstone mode $\omega^2 \sim \vq^2$ is not present in our DMFT 
calculation for the local charge susceptibility. The visible charge 
gap\cite{extrchis} $\Delta_{\rm c}$, which is always   
smaller than the spin gap, first increases with $U$ but at strong
coupling decreases again (see Fig. \ref{ch_gap_Udep_x0.5}). The maximum occurs
at a value, which is a bit larger than $U_c=2$ for the two-body bound state
problem. 

\noindent
One way of understanding this behavior is to consider the mapping of the
attractive Hubbard model to the half filled repulsive model in a magnetic
field. At strong coupling this can be mapped to a Heisenberg model with spin
coupling $J=4t^2/U$. Charge excitations in the attractive model correspond to
spin excitations in the repulsive model, and the characteristic scale for the
latter is $J$.  The dashed line in Fig. \ref{ch_gap_Udep_x0.5} proportional to
$4t^2/U$ with an adjusted prefactor reflects this dependence.

\paragraph*{Perspective -}
Let us comment on the range of applications of the extended DMFT-NRG
method. Within this approach low temperature static and dynamic response
functions can be calculated.
In the experimental situation for the BCS-BEC 
crossover, the trapped fermionic gases constitute an inhomogeneous
system. With a larger numerical effort these 
situations can be studied with real space DMFT extensions as done for the Mott
transition \cite{HCR08}. In the  experiments it is difficult to cool the fermions
down to very low temperatures compared to the 
Fermi energy. At finite temperature, but still in the symmetry-broken phase, the
excitation gap is expected to be smaller compared to our $T=0$ case  and the
width of the quasiparticle excitation larger \cite{SPSBM96}. A calculation at
finite temperature would be desirable to see how well the general features 
described here are still visible in the whole crossover regime. 

Moreover, with the given method other models
accessible by DMFT can be investigated. This includes, for instance, systems
with a local electron phonon coupling, which are relevant to understand the
fulleride superconductors. Also periodic Anderson type models and occurring
superconducting phases are accessible. 

Apart from these approximation to lattice models with a local self-energy, our
method could also be used to study open problems for magnetic impurities in
superconductors \cite{BVZ06}. According to Anderson's theorem and the work of
Abrikosov and Gor'kov \cite{AGD63,tinkham} a finite concentration of
non-magnetic impurities does not alter an s-wave superconducting state, but
magnetic impurities suppress the order substantially. So far there are,
however, few reliable quantitative microscopic studies of the local effect of
a single impurity (magnetic or non-magnetic) on a superconductor. Since this
is a non-perturbative problem our self-consistent NRG calculation could be an
excellent method to tackle this problem. An attempt to investigate this based
on the NRG method without self-consistency  was made by Sakai et
al. \cite{SSSS93}. 

\paragraph*{Conclusions -}
We have applied an extended  DMFT-NRG method to calculate self-consistent
solutions with superconducting symmetry breaking in the attractive Hubbard
model. This method can access static and dynamic quantities for any
coupling strength and doping. 
We have focused on the zero temperature one particle spectrum and the local
dynamic spin and charge susceptibilities for quarter filling. It was
shown that the excitation energies are reduced substantially from their mean
field results due to fluctuations. At weak coupling there are well-defined 
fermionic quasiparticle excitations, but as displayed in Fig.
\ref{width_Udep_x1} the width of these excitation increases at intermediate
coupling, such that sharp quasiparticle excitations are suppressed. In addition, the
different behavior of the excitation gaps in the dynamic charge and spin
susceptibilities has been analyzed quantitatively. 
We also pointed out that this work is not only relevant for
the attractive Hubbard model, but the approach can be used to study
superconductivity in other models, as well as for a microscopic description of 
the effect of impurities on superconductors.

\noindent{\bf Acknowledgment}\par
\noindent
We wish to thank N. Dupuis, P. Jakubczyk, W. Metzner, P. Strack and A. Toschi
for helpful discussions, W. Koller and D. Meyer for their earlier
contributions to the development of the NRG programs, and P. Strack for
 critically reading the manuscript.

\bibliography{artikel,biblio1,footnote}

\begin{thebibliography}{50}
\expandafter\ifx\csname natexlab\endcsname\relax\def\natexlab#1{#1}\fi
\expandafter\ifx\csname bibnamefont\endcsname\relax
  \def\bibnamefont#1{#1}\fi
\expandafter\ifx\csname bibfnamefont\endcsname\relax
  \def\bibfnamefont#1{#1}\fi
\expandafter\ifx\csname citenamefont\endcsname\relax
  \def\citenamefont#1{#1}\fi
\expandafter\ifx\csname url\endcsname\relax
  \def\url#1{\texttt{#1}}\fi
\expandafter\ifx\csname urlprefix\endcsname\relax\def\urlprefix{URL }\fi
\providecommand{\bibinfo}[2]{#2}
\providecommand{\eprint}[2][]{\url{#2}}

\bibitem[{\citenamefont{Bardeen et~al.}(1957)\citenamefont{Bardeen, Cooper, and
  Schrieffer}}]{BCS57}
\bibinfo{author}{\bibfnamefont{J.}~\bibnamefont{Bardeen}},
  \bibinfo{author}{\bibfnamefont{L.}~\bibnamefont{Cooper}}, \bibnamefont{and}
  \bibinfo{author}{\bibfnamefont{J.}~\bibnamefont{Schrieffer}},
  \bibinfo{journal}{Phys. Rev.} \textbf{\bibinfo{volume}{108}},
  \bibinfo{pages}{1175} (\bibinfo{year}{1957}).

\bibitem[{\citenamefont{Greiner et~al.}(2003)\citenamefont{Greiner, Regal, and
  Jin}}]{GRJ03}
\bibinfo{author}{\bibfnamefont{M.}~\bibnamefont{Greiner}},
  \bibinfo{author}{\bibfnamefont{C.}~\bibnamefont{Regal}}, \bibnamefont{and}
  \bibinfo{author}{\bibfnamefont{D.}~\bibnamefont{Jin}},
  \bibinfo{journal}{Nature} \textbf{\bibinfo{volume}{426}},
  \bibinfo{pages}{537} (\bibinfo{year}{2003}).

\bibitem[{\citenamefont{Zwierlein et~al.}(2004)\citenamefont{Zwierlein, Stan,
  Schunck, Raupach, Kerman, and Ketterle}}]{ZSSRKK04}
\bibinfo{author}{\bibfnamefont{M.}~\bibnamefont{Zwierlein}},
  \bibinfo{author}{\bibfnamefont{C.~A.} \bibnamefont{Stan}},
  \bibinfo{author}{\bibfnamefont{C.~H.} \bibnamefont{Schunck}},
  \bibinfo{author}{\bibfnamefont{S.~M.~F.} \bibnamefont{Raupach}},
  \bibinfo{author}{\bibfnamefont{A.~J.} \bibnamefont{Kerman}},
  \bibnamefont{and} \bibinfo{author}{\bibfnamefont{W.}~\bibnamefont{Ketterle}},
  \bibinfo{journal}{Phys. Rev. Lett.} \textbf{\bibinfo{volume}{92}},
  \bibinfo{pages}{120403} (\bibinfo{year}{2004}).

\bibitem[{\citenamefont{Zwierlein et~al.}(2005)\citenamefont{Zwierlein,
  Abo-Shaeer, Shirotzek, Schunck, and Ketterle}}]{ZSSSK05}
\bibinfo{author}{\bibfnamefont{M.}~\bibnamefont{Zwierlein}},
  \bibinfo{author}{\bibfnamefont{J.}~\bibnamefont{Abo-Shaeer}},
  \bibinfo{author}{\bibfnamefont{A.}~\bibnamefont{Shirotzek}},
  \bibinfo{author}{\bibfnamefont{C.~H.} \bibnamefont{Schunck}},
  \bibnamefont{and} \bibinfo{author}{\bibfnamefont{W.}~\bibnamefont{Ketterle}},
  \bibinfo{journal}{Nature} \textbf{\bibinfo{volume}{435}},
  \bibinfo{pages}{1047} (\bibinfo{year}{2005}).

\bibitem[{\citenamefont{Bloch et~al.}(2008)\citenamefont{Bloch, Dalibard, and
  Zwerger}}]{BDZ08}
\bibinfo{author}{\bibfnamefont{I.}~\bibnamefont{Bloch}},
  \bibinfo{author}{\bibfnamefont{J.}~\bibnamefont{Dalibard}}, \bibnamefont{and}
  \bibinfo{author}{\bibfnamefont{W.}~\bibnamefont{Zwerger}},
  \bibinfo{journal}{Rev. Mod. Phys.} \textbf{\bibinfo{volume}{80}},
  \bibinfo{pages}{885} (\bibinfo{year}{2008}).

\bibitem[{\citenamefont{Eagles}(1969)}]{Eag69}
\bibinfo{author}{\bibfnamefont{D.~M.} \bibnamefont{Eagles}},
  \bibinfo{journal}{Phys. Rev.} \textbf{\bibinfo{volume}{186}},
  \bibinfo{pages}{456} (\bibinfo{year}{1969}).

\bibitem[{\citenamefont{Leggett}(1980)}]{Leg80}
\bibinfo{author}{\bibfnamefont{A.~J.} \bibnamefont{Leggett}}, in
  \emph{\bibinfo{booktitle}{Modern Trends in the Theory of Condensed Matter}},
  edited by \bibinfo{editor}{\bibfnamefont{A.}~\bibnamefont{Pekalski}}
  \bibnamefont{and} \bibinfo{editor}{\bibfnamefont{R.}~\bibnamefont{Przystawa}}
  (\bibinfo{publisher}{Springer}, \bibinfo{address}{Berlin},
  \bibinfo{year}{1980}).

\bibitem[{\citenamefont{Nozi\`eres and Schmitt-Rink}(1985)}]{NS85}
\bibinfo{author}{\bibfnamefont{P.}~\bibnamefont{Nozi\`eres}} \bibnamefont{and}
  \bibinfo{author}{\bibfnamefont{S.}~\bibnamefont{Schmitt-Rink}},
  \bibinfo{journal}{J. Low Temp. Phys.} \textbf{\bibinfo{volume}{59}},
  \bibinfo{pages}{195} (\bibinfo{year}{1985}).

\bibitem[{\citenamefont{Randeria}(1995)}]{Ran95}
\bibinfo{author}{\bibfnamefont{M.}~\bibnamefont{Randeria}}, in
  \emph{\bibinfo{booktitle}{Bose-{E}instein {C}ondensation}}, edited by
  \bibinfo{editor}{\bibfnamefont{A.}~\bibnamefont{Griffin}},
  \bibinfo{editor}{\bibfnamefont{D.}~\bibnamefont{Snoke}}, \bibnamefont{and}
  \bibinfo{editor}{\bibfnamefont{S.}~\bibnamefont{Strinagari}}
  (\bibinfo{publisher}{Cambridge University Press},
  \bibinfo{address}{Cambridge}, \bibinfo{year}{1995}).

\bibitem[{\citenamefont{Micnas et~al.}(1990)\citenamefont{Micnas, Ranninger,
  and S.Robaszkiewicz}}]{MRR90}
\bibinfo{author}{\bibfnamefont{R.}~\bibnamefont{Micnas}},
  \bibinfo{author}{\bibfnamefont{J.}~\bibnamefont{Ranninger}},
  \bibnamefont{and} \bibinfo{author}{\bibnamefont{S.Robaszkiewicz}},
  \bibinfo{journal}{Rev. Mod. Phys.} \textbf{\bibinfo{volume}{62}},
  \bibinfo{pages}{113} (\bibinfo{year}{1990}).

\bibitem[{\citenamefont{Toschi et~al.}(2005{\natexlab{a}})\citenamefont{Toschi,
  Capone, and Castellani}}]{TCC05}
\bibinfo{author}{\bibfnamefont{A.}~\bibnamefont{Toschi}},
  \bibinfo{author}{\bibfnamefont{M.}~\bibnamefont{Capone}}, \bibnamefont{and}
  \bibinfo{author}{\bibfnamefont{C.}~\bibnamefont{Castellani}},
  \bibinfo{journal}{Phys. Rev. B} \textbf{\bibinfo{volume}{72}},
  \bibinfo{pages}{235118} (\bibinfo{year}{2005}{\natexlab{a}}).

\bibitem[{\citenamefont{Chen et~al.}(2006)\citenamefont{Chen, Levin, and
  Stajic}}]{CLS06}
\bibinfo{author}{\bibfnamefont{Q.}~\bibnamefont{Chen}},
  \bibinfo{author}{\bibfnamefont{K.}~\bibnamefont{Levin}}, \bibnamefont{and}
  \bibinfo{author}{\bibfnamefont{J.}~\bibnamefont{Stajic}},
  \bibinfo{journal}{J. Low Temp. Phys.} \textbf{\bibinfo{volume}{32}},
  \bibinfo{pages}{406} (\bibinfo{year}{2006}).

\bibitem[{\citenamefont{Chin et~al.}(2004)\citenamefont{Chin, Bartenstein,
  Altmeyer, Riedl, Jochim, {Hecker Denschlag}, and Grimm}}]{CBARJDG04}
\bibinfo{author}{\bibfnamefont{C.}~\bibnamefont{Chin}},
  \bibinfo{author}{\bibfnamefont{M.}~\bibnamefont{Bartenstein}},
  \bibinfo{author}{\bibfnamefont{A.}~\bibnamefont{Altmeyer}},
  \bibinfo{author}{\bibfnamefont{S.}~\bibnamefont{Riedl}},
  \bibinfo{author}{\bibfnamefont{S.}~\bibnamefont{Jochim}},
  \bibinfo{author}{\bibfnamefont{J.}~\bibnamefont{{Hecker Denschlag}}},
  \bibnamefont{and} \bibinfo{author}{\bibfnamefont{R.}~\bibnamefont{Grimm}},
  \bibinfo{journal}{Science} \textbf{\bibinfo{volume}{305}},
  \bibinfo{pages}{1128} (\bibinfo{year}{2004}).

\bibitem[{\citenamefont{Greiner et~al.}(2005)\citenamefont{Greiner, Regal, and
  Jin}}]{GRJ05}
\bibinfo{author}{\bibfnamefont{M.}~\bibnamefont{Greiner}},
  \bibinfo{author}{\bibfnamefont{C.~A.} \bibnamefont{Regal}}, \bibnamefont{and}
  \bibinfo{author}{\bibfnamefont{D.~S.} \bibnamefont{Jin}},
  \bibinfo{journal}{Phys. Rev. Lett.} \textbf{\bibinfo{volume}{94}},
  \bibinfo{pages}{070403} (\bibinfo{year}{2005}).

\bibitem[{\citenamefont{Keller et~al.}(2001)\citenamefont{Keller, Metzner, and
  Schollw\"ock}}]{KMS01}
\bibinfo{author}{\bibfnamefont{M.}~\bibnamefont{Keller}},
  \bibinfo{author}{\bibfnamefont{W.}~\bibnamefont{Metzner}}, \bibnamefont{and}
  \bibinfo{author}{\bibfnamefont{U.}~\bibnamefont{Schollw\"ock}},
  \bibinfo{journal}{Phys. Rev. Lett.} \textbf{\bibinfo{volume}{86}},
  \bibinfo{pages}{4612} (\bibinfo{year}{2001}).

\bibitem[{\citenamefont{Capone et~al.}(2002)\citenamefont{Capone, Castellani,
  and Grilli}}]{CCG02}
\bibinfo{author}{\bibfnamefont{M.}~\bibnamefont{Capone}},
  \bibinfo{author}{\bibfnamefont{C.}~\bibnamefont{Castellani}},
  \bibnamefont{and} \bibinfo{author}{\bibfnamefont{M.}~\bibnamefont{Grilli}},
  \bibinfo{journal}{Phys. Rev. Lett.} \textbf{\bibinfo{volume}{88}},
  \bibinfo{pages}{126403} (\bibinfo{year}{2002}).

\bibitem[{\citenamefont{Garg et~al.}(2005)\citenamefont{Garg, Krishnamurthy,
  and Randeria}}]{GKR05}
\bibinfo{author}{\bibfnamefont{A.}~\bibnamefont{Garg}},
  \bibinfo{author}{\bibfnamefont{H.~R.} \bibnamefont{Krishnamurthy}},
  \bibnamefont{and} \bibinfo{author}{\bibfnamefont{M.}~\bibnamefont{Randeria}},
  \bibinfo{journal}{Phys. Rev. B} \textbf{\bibinfo{volume}{72}},
  \bibinfo{pages}{024517} (\bibinfo{year}{2005}).

\bibitem[{\citenamefont{Toschi et~al.}(2005{\natexlab{b}})\citenamefont{Toschi,
  Barone, Capone, and Castellani}}]{TBCC05}
\bibinfo{author}{\bibfnamefont{A.}~\bibnamefont{Toschi}},
  \bibinfo{author}{\bibfnamefont{P.}~\bibnamefont{Barone}},
  \bibinfo{author}{\bibfnamefont{M.}~\bibnamefont{Capone}}, \bibnamefont{and}
  \bibinfo{author}{\bibfnamefont{C.}~\bibnamefont{Castellani}},
  \bibinfo{journal}{New J. Phys.} \textbf{\bibinfo{volume}{7}},
  \bibinfo{pages}{7} (\bibinfo{year}{2005}{\natexlab{b}}).

\bibitem[{\citenamefont{Singer et~al.}(1996)\citenamefont{Singer, Pedersen,
  Schneider, Beck, and Matuttis}}]{SPSBM96}
\bibinfo{author}{\bibfnamefont{J.~M.} \bibnamefont{Singer}},
  \bibinfo{author}{\bibfnamefont{M.~H.} \bibnamefont{Pedersen}},
  \bibinfo{author}{\bibfnamefont{T.}~\bibnamefont{Schneider}},
  \bibinfo{author}{\bibfnamefont{H.}~\bibnamefont{Beck}}, \bibnamefont{and}
  \bibinfo{author}{\bibfnamefont{H.-G.} \bibnamefont{Matuttis}},
  \bibinfo{journal}{Phys. Rev. B} \textbf{\bibinfo{volume}{54}},
  \bibinfo{pages}{1286} (\bibinfo{year}{1996}).

\bibitem[{\citenamefont{Kyung et~al.}(2006)\citenamefont{Kyung, Georges, and
  Tremblay}}]{KGT06}
\bibinfo{author}{\bibfnamefont{B.}~\bibnamefont{Kyung}},
  \bibinfo{author}{\bibfnamefont{A.}~\bibnamefont{Georges}}, \bibnamefont{and}
  \bibinfo{author}{\bibfnamefont{A.-M.~S.} \bibnamefont{Tremblay}},
  \bibinfo{journal}{Phys. Rev. B} \textbf{\bibinfo{volume}{74}},
  \bibinfo{pages}{024501} (\bibinfo{year}{2006}).

\bibitem[{\citenamefont{Wilson}(1975)}]{Wil75}
\bibinfo{author}{\bibfnamefont{K.}~\bibnamefont{Wilson}},
  \bibinfo{journal}{Rev. Mod. Phys.} \textbf{\bibinfo{volume}{47}},
  \bibinfo{pages}{773} (\bibinfo{year}{1975}).

\bibitem[{\citenamefont{Hewson}(1993)}]{hewson}
\bibinfo{author}{\bibfnamefont{A.~C.} \bibnamefont{Hewson}},
  \emph{\bibinfo{title}{The Kondo Problem to Heavy Fermions}}
  (\bibinfo{publisher}{Cambridge University Press},
  \bibinfo{address}{Cambridge}, \bibinfo{year}{1993}).

\bibitem[{\citenamefont{Sakai et~al.}(1989)\citenamefont{Sakai, Shimizu, and
  Kasuya}}]{SSK89}
\bibinfo{author}{\bibfnamefont{O.}~\bibnamefont{Sakai}},
  \bibinfo{author}{\bibfnamefont{Y.}~\bibnamefont{Shimizu}}, \bibnamefont{and}
  \bibinfo{author}{\bibfnamefont{T.}~\bibnamefont{Kasuya}},
  \bibinfo{journal}{J. Phys. Soc. Japan} \textbf{\bibinfo{volume}{58}},
  \bibinfo{pages}{3666} (\bibinfo{year}{1989}).

\bibitem[{\citenamefont{Costi et~al.}(1994)\citenamefont{Costi, Hewson, and
  Zlati{\'c}}}]{CHZ94}
\bibinfo{author}{\bibfnamefont{T.~A.} \bibnamefont{Costi}},
  \bibinfo{author}{\bibfnamefont{A.~C.} \bibnamefont{Hewson}},
  \bibnamefont{and}
  \bibinfo{author}{\bibfnamefont{V.}~\bibnamefont{Zlati{\'c}}},
  \bibinfo{journal}{J. Phys.: Cond. Mat.} \textbf{\bibinfo{volume}{6}},
  \bibinfo{pages}{2519} (\bibinfo{year}{1994}).

\bibitem[{\citenamefont{Georges et~al.}(1996)\citenamefont{Georges, Kotliar,
  Krauth, and Rozenberg}}]{GKKR96}
\bibinfo{author}{\bibfnamefont{A.}~\bibnamefont{Georges}},
  \bibinfo{author}{\bibfnamefont{G.}~\bibnamefont{Kotliar}},
  \bibinfo{author}{\bibfnamefont{W.}~\bibnamefont{Krauth}}, \bibnamefont{and}
  \bibinfo{author}{\bibfnamefont{M.}~\bibnamefont{Rozenberg}},
  \bibinfo{journal}{Rev. Mod. Phys.} \textbf{\bibinfo{volume}{68}},
  \bibinfo{pages}{13} (\bibinfo{year}{1996}).

\bibitem[{\citenamefont{Bulla et~al.}(1997)\citenamefont{Bulla, Pruschke, and
  Hewson}}]{BPH97}
\bibinfo{author}{\bibfnamefont{R.}~\bibnamefont{Bulla}},
  \bibinfo{author}{\bibfnamefont{T.}~\bibnamefont{Pruschke}}, \bibnamefont{and}
  \bibinfo{author}{\bibfnamefont{A.~C.} \bibnamefont{Hewson}},
  \bibinfo{journal}{J. Phys.: Cond. Mat.} \textbf{\bibinfo{volume}{9}},
  \bibinfo{pages}{10463} (\bibinfo{year}{1997}).

\bibitem[{\citenamefont{Hofstetter}(2000)}]{Hof00}
\bibinfo{author}{\bibfnamefont{W.}~\bibnamefont{Hofstetter}},
  \bibinfo{journal}{Phys. Rev. Lett.} \textbf{\bibinfo{volume}{85}},
  \bibinfo{pages}{1508} (\bibinfo{year}{2000}).

\bibitem[{\citenamefont{Anders and Schiller}(2005)}]{AS05}
\bibinfo{author}{\bibfnamefont{F.~B.} \bibnamefont{Anders}} \bibnamefont{and}
  \bibinfo{author}{\bibfnamefont{A.}~\bibnamefont{Schiller}},
  \bibinfo{journal}{Phys. Rev. Lett.} \textbf{\bibinfo{volume}{95}},
  \bibinfo{pages}{196801} (\bibinfo{year}{2005}).

\bibitem[{\citenamefont{Peters et~al.}(2006)\citenamefont{Peters, Pruschke, and
  Anders}}]{PPA06}
\bibinfo{author}{\bibfnamefont{R.}~\bibnamefont{Peters}},
  \bibinfo{author}{\bibfnamefont{T.}~\bibnamefont{Pruschke}}, \bibnamefont{and}
  \bibinfo{author}{\bibfnamefont{F.~B.} \bibnamefont{Anders}},
  \bibinfo{journal}{Phys. Rev. B} \textbf{\bibinfo{volume}{74}},
  \bibinfo{pages}{245114} (\bibinfo{year}{2006}).

\bibitem[{\citenamefont{Weichselbaum and von Delft}(2007)}]{WD07}
\bibinfo{author}{\bibfnamefont{A.}~\bibnamefont{Weichselbaum}}
  \bibnamefont{and} \bibinfo{author}{\bibfnamefont{J.}~\bibnamefont{von
  Delft}}, \bibinfo{journal}{Phys. Rev. Lett.} \textbf{\bibinfo{volume}{99}},
  \bibinfo{eid}{076402} (\bibinfo{year}{2007}).

\bibitem[{\citenamefont{Bulla et~al.}(2008)\citenamefont{Bulla, Costi, and
  Pruschke}}]{BCP08}
\bibinfo{author}{\bibfnamefont{R.}~\bibnamefont{Bulla}},
  \bibinfo{author}{\bibfnamefont{T.}~\bibnamefont{Costi}}, \bibnamefont{and}
  \bibinfo{author}{\bibfnamefont{T.}~\bibnamefont{Pruschke}},
  \bibinfo{journal}{Rev. Mod. Phys.} \textbf{\bibinfo{volume}{80}},
  \bibinfo{pages}{395} (\bibinfo{year}{2008}).

\bibitem[{\citenamefont{Bulla}(1999)}]{Bul99}
\bibinfo{author}{\bibfnamefont{R.}~\bibnamefont{Bulla}},
  \bibinfo{journal}{Phys. Rev. Lett.} \textbf{\bibinfo{volume}{83}},
  \bibinfo{pages}{136} (\bibinfo{year}{1999}).

\bibitem[{\citenamefont{Zitzler et~al.}(2002)\citenamefont{Zitzler, Pruschke,
  and Bulla}}]{ZPB02}
\bibinfo{author}{\bibfnamefont{R.}~\bibnamefont{Zitzler}},
  \bibinfo{author}{\bibfnamefont{T.}~\bibnamefont{Pruschke}}, \bibnamefont{and}
  \bibinfo{author}{\bibfnamefont{R.}~\bibnamefont{Bulla}},
  \bibinfo{journal}{Eur. Phys. J. B} \textbf{\bibinfo{volume}{27}},
  \bibinfo{pages}{473} (\bibinfo{year}{2002}).

\bibitem[{\citenamefont{Bauer and Hewson}(2007)}]{BH07c}
\bibinfo{author}{\bibfnamefont{J.}~\bibnamefont{Bauer}} \bibnamefont{and}
  \bibinfo{author}{\bibfnamefont{A.~C.} \bibnamefont{Hewson}},
  \bibinfo{journal}{Eur. Phys. J. B} \textbf{\bibinfo{volume}{57}},
  \bibinfo{pages}{235} (\bibinfo{year}{2007}).

\bibitem[{\citenamefont{Satori et~al.}(1992)\citenamefont{Satori, Shiba, Sakai,
  and Shimizu}}]{SSSS92}
\bibinfo{author}{\bibfnamefont{K.}~\bibnamefont{Satori}},
  \bibinfo{author}{\bibfnamefont{H.}~\bibnamefont{Shiba}},
  \bibinfo{author}{\bibfnamefont{O.}~\bibnamefont{Sakai}}, \bibnamefont{and}
  \bibinfo{author}{\bibfnamefont{Y.}~\bibnamefont{Shimizu}},
  \bibinfo{journal}{J. Phys. Soc. Japan} \textbf{\bibinfo{volume}{61}},
  \bibinfo{pages}{3239} (\bibinfo{year}{1992}).

\bibitem[{\citenamefont{Sakai et~al.}(1993)\citenamefont{Sakai, Shimizu, Shiba,
  and Satori}}]{SSSS93}
\bibinfo{author}{\bibfnamefont{O.}~\bibnamefont{Sakai}},
  \bibinfo{author}{\bibfnamefont{Y.}~\bibnamefont{Shimizu}},
  \bibinfo{author}{\bibfnamefont{H.}~\bibnamefont{Shiba}}, \bibnamefont{and}
  \bibinfo{author}{\bibfnamefont{K.}~\bibnamefont{Satori}},
  \bibinfo{journal}{J. Phys. Soc. Japan} \textbf{\bibinfo{volume}{62}},
  \bibinfo{pages}{3181} (\bibinfo{year}{1993}).

\bibitem[{\citenamefont{Bauer et~al.}(2007)\citenamefont{Bauer, Oguri, and
  Hewson}}]{BOH07}
\bibinfo{author}{\bibfnamefont{J.}~\bibnamefont{Bauer}},
  \bibinfo{author}{\bibfnamefont{A.}~\bibnamefont{Oguri}}, \bibnamefont{and}
  \bibinfo{author}{\bibfnamefont{A.}~\bibnamefont{Hewson}},
  \bibinfo{journal}{J. Phys.: Cond. Mat.} \textbf{\bibinfo{volume}{19}},
  \bibinfo{pages}{486211} (\bibinfo{year}{2007}).

\bibitem[{\citenamefont{Hecht et~al.}(2008)\citenamefont{Hecht, Weichselbaum,
  {von Delft}, and Bulla}}]{HWDB08}
\bibinfo{author}{\bibfnamefont{T.}~\bibnamefont{Hecht}},
  \bibinfo{author}{\bibfnamefont{A.}~\bibnamefont{Weichselbaum}},
  \bibinfo{author}{\bibfnamefont{J.}~\bibnamefont{{von Delft}}},
  \bibnamefont{and} \bibinfo{author}{\bibfnamefont{R.}~\bibnamefont{Bulla}},
  \bibinfo{journal}{J. Phys.: Cond. Mat.} \textbf{\bibinfo{volume}{20}},
  \bibinfo{pages}{275213} (\bibinfo{year}{2008}).

\bibitem[{\citenamefont{Yoshioka and Ohashi}(2000)}]{YO00}
\bibinfo{author}{\bibfnamefont{T.}~\bibnamefont{Yoshioka}} \bibnamefont{and}
  \bibinfo{author}{\bibfnamefont{Y.}~\bibnamefont{Ohashi}},
  \bibinfo{journal}{J. Phys. Soc. Japan} \textbf{\bibinfo{volume}{69}},
  \bibinfo{pages}{1812} (\bibinfo{year}{2000}).

\bibitem[{\citenamefont{Choi et~al.}(2004)\citenamefont{Choi, Lee, Kang, and
  Belzig}}]{CLKB04}
\bibinfo{author}{\bibfnamefont{M.-S.} \bibnamefont{Choi}},
  \bibinfo{author}{\bibfnamefont{M.}~\bibnamefont{Lee}},
  \bibinfo{author}{\bibfnamefont{K.}~\bibnamefont{Kang}}, \bibnamefont{and}
  \bibinfo{author}{\bibfnamefont{W.}~\bibnamefont{Belzig}},
  \bibinfo{journal}{Phys. Rev. B} \textbf{\bibinfo{volume}{70}},
  \bibinfo{pages}{020502} (\bibinfo{year}{2004}).

\bibitem[{\citenamefont{Oguri et~al.}(2004)\citenamefont{Oguri, Tanaka, and
  Hewson}}]{OTH04}
\bibinfo{author}{\bibfnamefont{A.}~\bibnamefont{Oguri}},
  \bibinfo{author}{\bibfnamefont{Y.}~\bibnamefont{Tanaka}}, \bibnamefont{and}
  \bibinfo{author}{\bibfnamefont{A.~C.} \bibnamefont{Hewson}},
  \bibinfo{journal}{J. Phys. Soc. Japan} \textbf{\bibinfo{volume}{73}},
  \bibinfo{pages}{2496} (\bibinfo{year}{2004}).

\bibitem[{\citenamefont{Bauer et~al.}(2009)\citenamefont{Bauer, Hewson, and
  Dupuis}}]{BHD09pre}
\bibinfo{author}{\bibfnamefont{J.}~\bibnamefont{Bauer}},
  \bibinfo{author}{\bibfnamefont{A.~C.} \bibnamefont{Hewson}},
  \bibnamefont{and} \bibinfo{author}{\bibfnamefont{N.}~\bibnamefont{Dupuis}}
  (\bibinfo{year}{2009}), \bibinfo{note}{cond-mat/0901.1760}.

\bibitem[{exp()}]{expansion}
\bibinfo{note}{The following expansion is justified, when the imaginary parts
  of the self-energies are small and do not vary much near $E_{\vk}$, which is
  satisfied in the weak coupling limit. However, as seen in Fig.
  \ref{width_Udep_x1} and Ref. \onlinecite{BHD08tbp} such an analysis breaks
  down in the intermediate coupling regime. It would therefore be misleading to
  use the results for strong coupling.}

\bibitem[{\citenamefont{Martin-Rodero and Flores}(1992)}]{MF92}
\bibinfo{author}{\bibfnamefont{A.}~\bibnamefont{Martin-Rodero}}
  \bibnamefont{and} \bibinfo{author}{\bibfnamefont{F.}~\bibnamefont{Flores}},
  \bibinfo{journal}{Phys. Rev. B} \textbf{\bibinfo{volume}{45}},
  \bibinfo{pages}{13008} (\bibinfo{year}{1992}).

\bibitem[{ext()}]{extrchis}
\bibinfo{note}{The gap in the dynamic response functions was extracted by
  taking the value of $\omega$ where the imaginary part of $\chi$ has increased
  from zero to $5\%$ of its maximal value.}

\bibitem[{\citenamefont{Helmes et~al.}(2008)\citenamefont{Helmes, Costi, and
  Rosch}}]{HCR08}
\bibinfo{author}{\bibfnamefont{R.~W.} \bibnamefont{Helmes}},
  \bibinfo{author}{\bibfnamefont{T.~A.} \bibnamefont{Costi}}, \bibnamefont{and}
  \bibinfo{author}{\bibfnamefont{A.}~\bibnamefont{Rosch}},
  \bibinfo{journal}{Phys. Rev. Lett.} \textbf{\bibinfo{volume}{100}},
  \bibinfo{eid}{056403} (\bibinfo{year}{2008}).

\bibitem[{\citenamefont{Balatsky et~al.}(2006)\citenamefont{Balatsky, Vekhter,
  and Zhu}}]{BVZ06}
\bibinfo{author}{\bibfnamefont{A.~V.} \bibnamefont{Balatsky}},
  \bibinfo{author}{\bibfnamefont{I.}~\bibnamefont{Vekhter}}, \bibnamefont{and}
  \bibinfo{author}{\bibfnamefont{J.-X.} \bibnamefont{Zhu}},
  \bibinfo{journal}{Rev. Mod. Phys.} \textbf{\bibinfo{volume}{78}},
  \bibinfo{eid}{373} (\bibinfo{year}{2006}).

\bibitem[{\citenamefont{Abrikosov et~al.}(1963)\citenamefont{Abrikosov, Gorkov,
  and Dzyaloshinski}}]{AGD63}
\bibinfo{author}{\bibfnamefont{A.~A.} \bibnamefont{Abrikosov}},
  \bibinfo{author}{\bibfnamefont{L.~P.} \bibnamefont{Gorkov}},
  \bibnamefont{and} \bibinfo{author}{\bibfnamefont{I.~E.}
  \bibnamefont{Dzyaloshinski}}, \emph{\bibinfo{title}{Methods of quantum field
  theory in statistical physics}} (\bibinfo{publisher}{Dover},
  \bibinfo{address}{New York}, \bibinfo{year}{1963}).

\bibitem[{\citenamefont{Tinkham}(1996)}]{tinkham}
\bibinfo{author}{\bibfnamefont{M.}~\bibnamefont{Tinkham}},
  \emph{\bibinfo{title}{{I}ntroduction to {S}uperconductivity (2nd edition)}}
  (\bibinfo{publisher}{Mc Grawth-Hill Inc.}, \bibinfo{year}{1996}).

\bibitem[{BHD()}]{BHD08tbp}
\bibinfo{note}{J. Bauer, A.C. Hewson, and N. Dupuis, preprint in preparation
  (2008).}

\end{thebibliography}

\end{document}